\documentclass[preprint,5p,times,twocolumn]{elsarticle}

\usepackage{amssymb}
\usepackage{amsmath}

\usepackage{booktabs}
\usepackage{multirow}
\usepackage{subcaption}

\journal{Pattern Recognition Letters}

\begin{document}

\begin{frontmatter}



\title{Similarity-Based Supervised User Session Segmentation Method for Behavior Logs}


\author[1]{Yongzhi Jin}
\author[1]{Kazushi Okamoto}
\author[1]{Kei Harada}
\author[1]{Atsushi Shibata}
\author[1]{Koki Karube}

\affiliation[1]{organization={The University of Electro-Communications},
            addressline={1-5-1 Chofugaoka}, 
            city={Chofu},
            postcode={182-8585}, 
            state={Tokyo},
            country={Japan}}

\begin{abstract}
In information recommendation, a session refers to a sequence of user actions within a specific time frame. 
Session-based recommender systems aim to capture short-term preferences and generate relevant recommendations.
However, user interests may shift even within a session, making appropriate segmentation essential for modeling dynamic behaviors.
In this study, we propose a supervised session segmentation method based on similarity features derived from action embeddings and attributes.
We compute the similarity scores between items within a fixed-size window around each candidate segmentation point, using item co-occurrence embeddings, text embeddings of titles and brands, and price information as sources for these similarity features.
These features are used to train supervised classification models to predict the session boundaries.
We construct a manually annotated dataset from real browsing histories and evaluate the segmentation performance using F1-score, PR-AUC, and ROC-AUC.
The LightGBM model achieves the best performance, with an F1-score of 0.806 and a PR-AUC of 0.831.
These results demonstrate the effectiveness of the proposed method for session segmentation and its potential to capture dynamic user behaviors.
\end{abstract}




\end{frontmatter}



\section{Introduction}
Recommender systems (RSs) have been increasingly adopted across various online service domains, including video, music, and e-commerce platforms.
In this regard, there has been an increasing focus on the algorithms behind RSs, with session-based RSs (SBRSs) being particularly effective in capturing recent user preferences~\cite{2021_S.Wang}.
A session refers to a sequence of user actions performed within a certain time frame while using an application or service.
For instance, a user's browsing history is a typical session.
An SBRS captures user preferences from session data and generates relevant item recommendations.
However, user preferences are dynamic, and sessions often contain multiple topics.
Using entire sessions for training may degrade prediction accuracy.
Although session segmentation is crucial for the effective processing of session data, research on segmentation methods is limited.

Recently, several approaches have been explored for session segmentation.
Zhang et al.~\cite{2020_Y.Zhang} introduced a method based on cosine similarity between item embeddings.
Inspired by this idea, we conducted a unified evaluation of session segmentation methods, each constructed by combining behavioral and textual embeddings with supervised models~\cite{2024_Y.Jin}.
However, these methods have several limitations.
They fail to fully exploit the rich information available in the dataset and rely solely on local features to determine session boundaries, ignoring broader contexts such as features derived from a larger window size.

In this study, we propose a supervised session segmentation method that uses similarity-based features derived from item embeddings and additional attributes.
Each potential segmentation point is evaluated by computing the similarity scores within a local window centered on that point.
In addition, item-level information, such as price, is incorporated to enrich the feature representation.
These features are then used in a supervised learning framework to predict whether each point is a session boundary.
Specifically, in addition to the cosine similarities from various embeddings, we define a custom similarity metric based on item prices and include it as a feature.
Motivated by the above, we address the following research questions (RQs):
\begin{itemize}
  \setlength{\leftskip}{2em}
  \item[\textbf{RQ1:}]{How effective is the proposed method in segmenting session data?}
  \item[\textbf{RQ2:}]{Which supervised learning model is most effective for session segmentation?}
\end{itemize}
To address the RQs, we use real browsing history data from Amazon.
We apply natural language processing (NLP)-based embedding techniques to each browsing item and segment user behaviors based on the similarity scores between item representations.
These similarities are used as features in machine learning (ML) models, including LightGBM~\cite{2017_G.Ke}, XGBoost~\cite{2016_T.Chen}, CatBoost~\cite{2018_L.Prokhorenkova}, support vector machine (SVM)~\cite{1995_C.Cortes}, and logistic regression (LR).
We conduct comparative experiments against existing baselines to evaluate the effectiveness of the proposed method.
Because the original dataset lacked segmentation labels, we recruited annotators to label behavioral change points and construct the ground-truth dataset.

\section{Related Work}
A session segmentation method aims to identify segmentation points in sequences of user-item interactions within a continuous time period in the e-commerce domain.
The concept of a session varies across domains.
For instance, in the web domain, sessions are typically represented as sequences of user queries or web page visits; in dialogue systems, they correspond to exchanges of utterances between users; and in the e-commerce domain, they are often defined by users' browsing histories.
This section provides a brief overview of how session segmentation has been approached in various domains, such as the web and dialogue systems, and then focuses on its application and challenges in the e-commerce domain.

\subsection{Session Segmentation for Web}
In the web domain, session segmentation plays a crucial role.
In search engine systems, dividing and analyzing user behavior at the session level facilitates a more accurate understanding of the search context.
This contextual information can be leveraged for user-oriented learning methods.
Furthermore, predicting users' next search queries by capturing topic transitions within sessions enables more effective suggestions.

One of the fundamental session segmentation methods is the time-interval-based approach~\cite{2000_He}.
Several clustering-based approaches have been proposed for session segmentation.
For instance, the COBWEB-based method~\cite{2012_Z.Hou} uses a search-engine dataset and employs features such as time intervals and query likelihoods to learn a session segmentation model in an unsupervised manner.
However, feature engineering in this approach remains basic and does not fully leverage the rich content information available in the data.

To address this limitation, a supervised method employing Gradient Boosting Decision Trees (GBDTs) was proposed to estimate the probability that two pages belong to the same session~\cite{2013_Y.Ustinovskiy}.
Their approach uses anonymized browsing logs collected from a major search engine via a browser toolbar, and leverages a rich set of features, including URL-based, textual, and temporal features.
Inspired by this work, we formulate the session segmentation task as a binary classification problem to predict whether a session boundary exists between consecutive items.
This formulation enables the use of various ML models, including GBDT such as LightGBM, XGBoost, and CatBoost, as well as classical models such as SVM and LR.

\subsection{Session Segmentation for Dialogue}
Semantic segmentation techniques have been applied to a variety of tasks beyond the RS domain, such as dialogue and chat analyses~\cite{2016_Y.Song,2024_S.Xiong, 2018_H.Lee,2023_H.Gao}. 
In particular, Lee et al.~\cite{2018_H.Lee} utilized Doc2Vec~\cite{2014_Q.Le} to encode user utterances within conversations. 
This approach enables session segmentation by detecting shifts in semantic similarity across consecutive utterances, allowing the system to identify topic transitions effectively.
For targeted advertising, it is essential to extract and classify utterance blocks from instant messages that reflect user interests.
Lee et al. performed dialogue session segmentation by vectorizing utterances using Doc2Vec and dividing sessions based on their semantic consistency, leveraging cosine similarity.
The approach demonstrated the effectiveness of embedding-based representations for identifying topic shifts, which are highly relevant to segmentation tasks across various domains, including the present study. 
Nevertheless, although most research has focused on dialogue or chat sessions, session segmentation methods in the e-commerce domain, particularly for tasks such as targeted advertising, remain largely unexplored.

\subsection{Session Segmentation for E-commerce}
In e-commerce, an accurate understanding of user search behaviors is crucial for delivering sophisticated recommendations.
Session-based behavioral analysis has been actively studied to better capture the user context~\cite{2020_Y.Ma,2022_J.Seol}.
However, session segmentation has typically been performed using heuristic time thresholds lacking a rigorous or data-driven basis.

An embedding approach to session segmentation was proposed~\cite{2020_Y.Zhang}.
The study developed a method that segmented user sessions, such as browsing histories or item ratings, into multiple dummy sessions using behavior-based embeddings and cosine similarity.
We extended session segmentation approaches by incorporating not only Item2Vec item embeddings but also embeddings of item titles, and conducted a comparative evaluation of supervised and unsupervised models~\cite{2024_Y.Jin}.
The results demonstrated that using the similarity between Item2Vec embeddings outperformed approaches that directly input embedding vectors into the model.
However, this approach has several limitations.
Specifically, it uses a narrow window size that considers only the immediate context around the segmentation point and relies solely on embeddings as features.
This design fails to fully capture the relationships between items.

This study extends our previous work~\cite{2024_Y.Jin} and aims to develop a more comprehensive session segmentation method by refining and expanding its approach.
Our earlier findings suggested that cosine similarity between item embeddings effectively identifies meaningful session boundaries. 
Based on this insight, we incorporate similarity scores between the current and neighboring items as features in the segmentation model.
\section{Proposed Model}
In this study, we propose a supervised session segmentation method that predicts whether a point within a sequence of user interactions marks a session boundary.
The approach leverages similarity-based features derived from item embeddings and attributes, and applies ML models to perform segmentation.
The proposed method consists of the following main components.

\subsection{Item Embeddings}
\subsubsection{Behavior-based Embedding}
First, we introduce an embedding approach based on Item2Vec~\cite{2016_O.Barkan}, which is derived from Word2Vec~\cite{2013_T.Mikolov}. 
Word2Vec is a neural embedding model originally developed for NLP tasks. 
It transforms words into dense vector representations based on their surrounding context and grounded in the distributional hypothesis.
Item2Vec adapts this principle to behavioral data in e-commerce by treating user sessions as sequences of interactions and learning item vector representations that reflect their contextual co-occurrence with other items.

In this study, we employ Item2Vec as a behavior-based embedding method to capture relationships between items from session-level co-occurrence patterns.
We train the model using the Word2Vec implementation in the Gensim library, where each session is treated as analogously to a sentence in NLP.

\subsubsection{Text-based Embedding}
We incorporate text-based embeddings of item titles and brand names to capture semantic relationships between items beyond user behavior.
Traditionally, text embeddings are obtained by tokenizing text, embedding each word using Word2Vec~\cite{2013_T.Mikolov}, and averaging the resulting word vectors.
This simple yet efficient approach has been widely adopted across various NLP tasks~\cite{2023_I.Yamada}.
However, such methods are limited in modeling complex contextual semantics.
In this study, we use the text-embedding-3-small model, a state-of-the-art text-embedding model developed by OpenAI.
This model produces high-quality vector representations of text and demonstrates improved performance over its predecessor, text-embedding-ada-002.
It achieves higher performance across various embedding tasks, particularly in multilingual and English-language benchmarks, making it well suited for applications such as text retrieval, similarity analysis, and semantic search.

\subsection{Point-wise Feature Extraction}
The proposed method computes the following four types of similarity scores around each segmentation point:
\begin{itemize}
  \item{{\em Behavior sim.}: cosine similarity between item embeddings obtained from the Item2Vec model.}
  \item{{\em Brand sim.}: cosine similarity between brand embeddings generated by text-embedding-3-small.}
  \item{{\em Title sim.}: cosine similarity between product title embeddings generated by text-embedding-3-small.}
  \item{{\em Price sim.}: a score derived from the ratio of item prices.}
\end{itemize}
We define similarity functions $S_{t}(u_{i}, u_{j})$, where $t \in \{$ $\text{behavior}$, $\text{brand}$, $\text{title}$, $\text{price} \}$, between two items $u_{i}$ and $u_{j}$ to compute the corresponding behavior, brand, title, and price similarities.
The functions $S_{\text{behavior}}$, $S_{\text{brand}}$, and $S_{\text{title}}$ are defined as
\begin{align*}
  S_{t}(u_{i}, u_{j}) &= \frac{\boldsymbol{v}_{u_{i}}^{T} \cdot \boldsymbol{v}_{u_{j}}}{\| \boldsymbol{v}_{u_{i}} \| \| \boldsymbol{v}_{u_{j}} \|} \quad u_{i}, u_{j} \in W
\end{align*}
where $W$ denotes the set of items within a window, and $\boldsymbol{v}$ denotes the behavior, brand, or title embedding of an item.
The price similarity $S_{\text{price}}$ is defined as
\begin{align*}
  S_{\text{price}}(u_{i}, u_{j}) &=\exp \left( - \frac{|p_{i} - p_{j}|}{\min(p_{i + 1}, p_{j + 1})} 
 \right) \quad u_{i}, u_{j} \in W
\end{align*}
where $p$ denotes the price of item $u$.

\begin{figure}[t]
  \centering
  \includegraphics[width=0.95\linewidth]{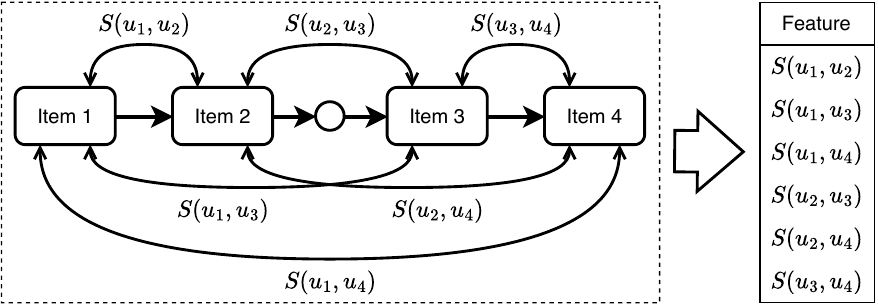}
  \caption{Example of feature extraction for window size $w = 2$}
  \label{fig:feature}
\end{figure}

As illustrated in Fig.~\ref{fig:feature}, a sliding window is applied around each candidate segmentation point to extract contextual information.
The window size $w$ determines the range of items considered.
If $w = 1$, only the immediately preceding and following items are included, whereas if $w = 4$, four items before and four after the segmentation point are included.
This results in $2w$ referenced items, yielding $d = {}_{2w}C_{2}$ pairwise comparisons.
All pairwise combinations within this window are used to compute similarity scores, which are aggregated into a feature vector $\boldsymbol{x} \in \mathbb{R}^{4d}$ representing the local context.

\subsection{Segmentation Point Classification}
After feature extraction, we apply a supervised classifier $f_{\boldsymbol{\theta}}: \mathbb{R}^{4d} \longrightarrow \{ 0, 1 \}$ to identify whether the input $\boldsymbol{x}$ is a segmentation point, where $\boldsymbol{\theta}$ denotes a learned parameter vector.
The parameter $\boldsymbol{\theta}$ is estimated using $n$ training samples $(y, \boldsymbol{x}) \in \{ 0, 1 \} \times \mathbb{R}^{4d}$.
Here, a label of 1 indicates a segmentation point, and 0 denotes a non-segmentation point.
In this study, we employ five classifiers: LightGBM, XGBoost, CatBoost, SVM, and LR.
\section{Session-Segmentation Experiment}
This experiment evaluated segmentation performance to examine the effectiveness of the proposed method (RQ1) and to compare the performance of different supervised learning models (RQ2).
This section outlines the dataset, annotation workflow, and evaluation procedure used in this study.

\begin{table}[t]
  \centering
  \caption{Statistical summary of session dataset (locale = JP)}
  \label{tab:session_statistics}
  \begin{tabular}{ll}
    \toprule
    Total Sessions & 979,119 \\
    Total Items & 389,888 \\
    Mean Session Length & 5.48 items \\
    Standard Deviation & 4.08 items \\
    Minimum Session Length & 3 items \\
    Median Session Length & 4 items \\
    Maximum Session Length & 475 items \\
    \bottomrule
  \end{tabular}
\end{table}
\begin{figure}[t]
  \centering
  \includegraphics[width=0.95\linewidth]{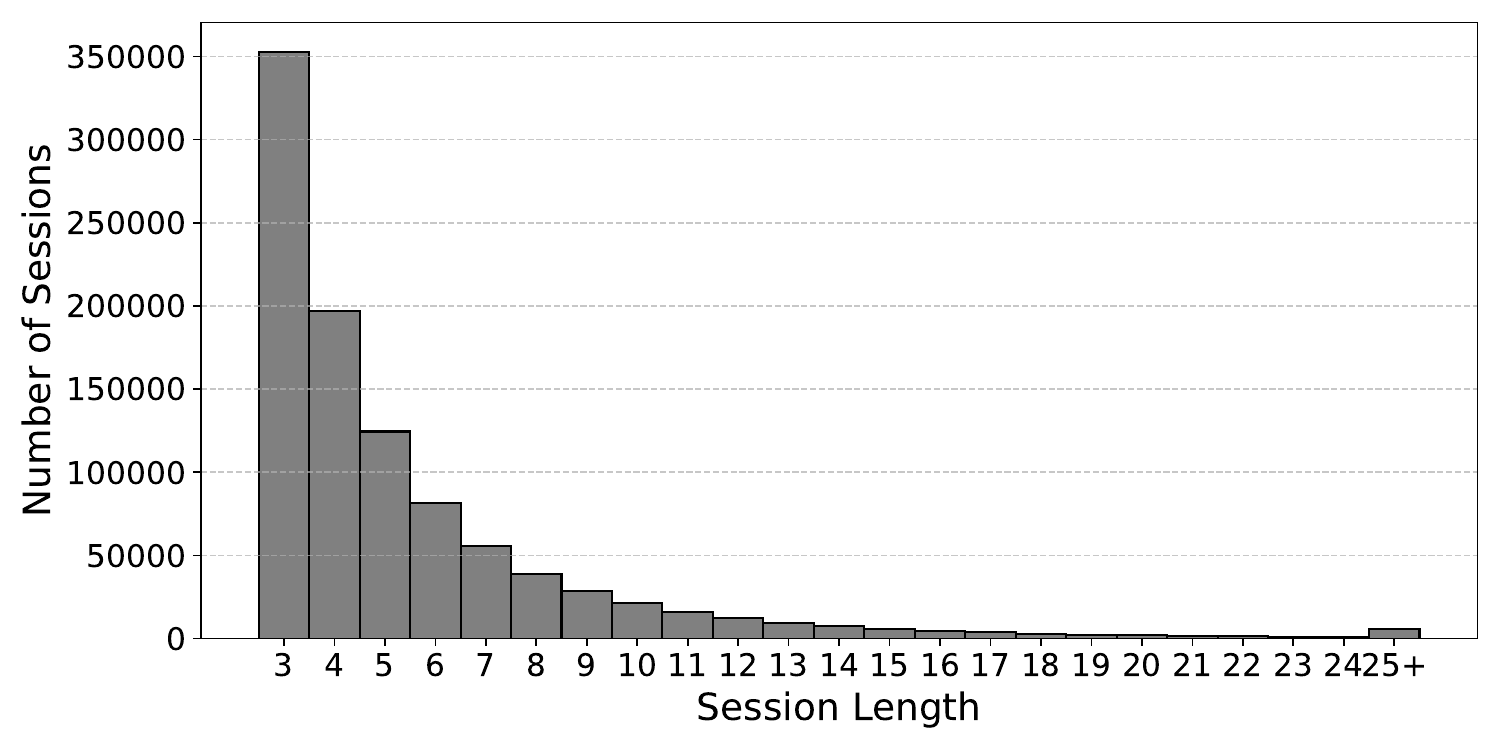}
  \caption{Distribution of session lengths}
  \label{fig:session_length_distribution}
\end{figure}

\subsection{Dataset}
In this experiment, we used the Amazon M2 dataset~\cite{2023_W.Jin}.
This dataset provides customer session data and is notable for covering six locales (English, German, Japanese, French, Italian, and Spanish) unlike conventional session datasets.
However, this study focuses exclusively on the Japanese data (locale = JP).
Each item within a session contains a locale, an ID, and several attributes, such as title, price, brand, color, size, model, material, author, and description.
Among these, we primarily utilized three representative attributes---title, price, and brand---as they capture the essential characteristics of an item.

The dataset comprises two components: {\em prev\_items}, representing items viewed by users, and {\em next\_item}, indicating items actually purchased.
These components are merged and treated as a single session.
Table~\ref{tab:session_statistics} presents a statistical summary of the session dataset with the Japanese locale (JP).
The distribution of session lengths is shown in Fig.~\ref{fig:session_length_distribution}. 
As illustrated, sessions of length three were the most frequent, and the frequency decreasing as the session length increased.

\subsection{Annotation}
Since the Amazon M2 dataset did not include segmentation point labels, we conducted an annotation process involving eight participants from the authors' laboratory.
Session data for annotation were randomly sampled from the full set of available sessions.
Consequently, a total of 2,400 sessions were annotated.
Among them, 10,904 segmentation points were identified, including 1,230 positive (segmentation) and 9,674 negative (non-segmentation) examples.
Hereafter, we define the annotated data such that segmentation points are labeled as 1 and non-segmentation points as 0.

Table~\ref{table:segmentation_point_distribution} summarizes the number of segmentation points labeled within the annotated sessions, as well as their distribution per session.
Each row corresponds to the segmentation results produced by a different annotator.
The results indicate that approximately 40\% of the sessions required segmentation. 
Furthermore, the small variation among annotators suggests that the annotations were consistent and reliable.

\begin{table}[t]
  \caption{Distribution of segmentation points labeled in each session by the annotator}
  \label{table:segmentation_point_distribution}
  \centering
  \begin{tabular}{ccccc}
  \toprule
  Annotator & 0 & 1 & 2 & 3+ \\
  \midrule
  1 & 0.577 & 0.287 & 0.097 & 0.040\\
  2 & 0.600 & 0.300 & 0.070 & 0.030\\ 
  3 & 0.650 & 0.273 & 0.063 & 0.013\\
  4 & 0.590 & 0.330 & 0.063 & 0.017\\
  5 & 0.573 & 0.333 & 0.053 & 0.040\\
  6 & 0.620 & 0.240 & 0.103 & 0.037\\
  7 & 0.640 & 0.287 & 0.060 & 0.013\\
  8 & 0.753 & 0.220 & 0.020 & 0.007\\
  \bottomrule
  \end{tabular}
\end{table}

Finally, we analyze the number of segmentation points per session. 
We defined sessions containing five or more items as {\em long sessions}, and those with fewer than five items as {\em short sessions}.
Table~\ref{table:segment_number_distribution} presents the distribution of the annotation results.
The first row corresponds to long sessions, whereas the second corresponds to short sessions.
Each column indicates the number of segmentation points within a session, and the numerical values show the number of sessions corresponding to each combination.
As shown in Table~\ref{table:segment_number_distribution}, long sessions tended to contain more segmentation points. 
The percentage of sessions requiring segmentation was 32.4\% for short sessions and 44.1\% for long sessions.
This result aligns with intuitive expectations.

\begin{table}[t]
  \caption{Distribution of segment numbers and session length types}
  \label{table:segment_number_distribution}
  \centering
  \begin{tabular}{ccccc}
  \toprule
  Length & 0 & 1 & 2 & 3+ \\
  \midrule
  Short session & 923 & 379 & 60 & 4 \\
  Long session & 578 & 302 & 99 & 55 \\
  \bottomrule
  \end{tabular}
\end{table}

\subsection{Evaluation Procedure}
\subsubsection{Overview of the Segmentation Task}
\begin{figure}[t]
  \centering
  \includegraphics[width=0.8\linewidth]{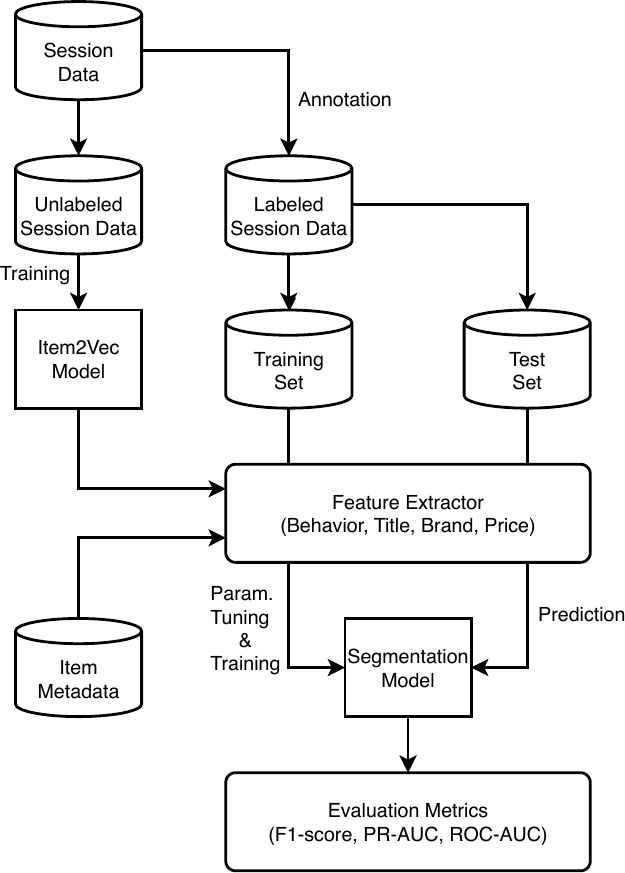}
  \caption{Experiment flow}
  \label{fig:experiment_flow}
\end{figure}

In this experiment, we address session segmentation as a binary classification task.
The objective is to identify whether a point in a sequence of user interactions corresponds to a session boundary.
The experimental pipeline is illustrated in Fig.~\ref{fig:experiment_flow}.

We used all session data summarized in Table~\ref{table:segmentation_point_distribution} to train an Item2Vec model, which generated item embeddings based on co-occurrence within sessions.
To prevent data leakage, annotated sessions were excluded from training, and only the remaining unlabeled sessions were used.
In parallel, item titles and brands stored in the database were embedded using the text-embedding-3-small model to capture semantic similarities.

For each candidate segmentation point in a session, four types of similarity scores, $S_{\text{behavior}}(u_{i}, u_{j})$, $S_{\text{brand}}(u_{i}, u_{j})$, $S_{\text{title}}(u_{i}, u_{j})$, and $S_{\text{price}}(u_{i}, u_{j})$, were computed for all item pairs $(u_{i}, u_{j})$ such that $u_{i}, u_{j} \in W$ and $u_{i} \ne u_{j}$, where $W$ denotes the set of items within a local window.
These scores were aggregated to construct input vectors for supervised classification.
The annotated segmentation labels were then used to train the classification models, which output a probability score indicating the likelihood that each point is a session boundary.

To assess segmentation performance, we used standard classification metrics: F1-score, the area under the precision-recall curve (PR-AUC), and the area under the receiver operating characteristic curve (ROC-AUC).
These metrics were used to evaluate the proposed method (RQ1) and to compare the performance of various classification models (RQ2).

\subsubsection{Data Partitioning and Hyperparameter Tuning}
The annotated dataset $D = \{ (y_{i}, \boldsymbol{x}_{i}) \}_{i = 1}^{n}$ was split into training and test sets, denoted as $D_{\text{train}}$ and $D_{\text{test}}$, respectively, with a 4:1 ratio.
In this experiment, we used $D_{\text{train}}$ to train the classification models, and evaluated their segmentation performance on $D_{\text{test}}$.
The hyperparameters of the models were optimized via five-fold cross-validation (CV) on $D_{\text{train}}$ using Optuna~\cite{2019_T.Akiba}.
The optimized hyperparameters were then used to evaluate the classification models on $D_{\text{test}}$.

A key consideration in this CV process was avoiding data leakage.
Input features were extracted from data points around segmentation boundaries.
However, if segmentation points from the same session appear in both the training and validation sets, the model may learn session-specific patterns rather than generalizable features, resulting in overestimated performance.
To mitigate this issue, we employed a Group K-Fold CV, treating each session as a distinct group.
Finally, we performed hyperparameter optimization, with the F1-score as the objective metric.

\subsubsection{Padding and Missing Value Handling}
We refer to the features of surrounding items when identifying whether a point corresponds to a segmentation boundary.
For each candidate point, a fixed-size window is applied to extract contextual items from both the preceding and following items.
However, if the window size extends beyond the session boundaries (e.g., at the beginning or end of a session), padding is applied by repeating the nearest item.
For instance, given a session [$u_{1}$, $u_{2}$, $u_{3}$, $u_{4}$] and a candidate point between $u_{1}$ and $u_{2}$, a window size of two results in the padded context [$u_{1}$, $u_{1}$, $u_{2}$, $u_{3}$], where the leftmost item is duplicated to maintain the required window size.
This design ensures that the input vector maintains a consistent dimensionality, regardless of the position of the candidate point within the session.

As described in Section 3.1, we trained the Item2Vec model using session data.
The annotated sessions were excluded from training to prevent potential data leakage.
This exclusion caused some items to appear only in the test set, lacking corresponding Item2Vec embeddings during testing.
Consequently, training models such as SVM and LR became difficult because similarity scores for such items could not be computed.
To mitigate this issue, we assigned a similarity score of zero where the embedding of either item was unavailable.

\section{Results and Discussion}
This section reports the session segmentation performance of both the baseline and proposed methods on the test set $D_{\text{test}}$.
Table~\ref{tab:performance} summarizes the evaluation results (F1-scores calculated using a threshold of 0.5).
In Table~\ref{tab:performance}, the columns represent the classification methods, whereas the rows are grouped by evaluation metrics, and present the results for different window sizes.
The highest score for each metric is highlighted in bold.

\begin{table*}[t]
  \centering
  \caption{Session segmentation performance}
  \label{tab:performance}
  \begin{tabular}{l c cccccc}
    \toprule
    Metric & $w$ & LightGBM & XGBoost & CatBoost & SVM & LR & Cosine Similarity \\
    \midrule
    \multirow{4}{*}{F1-score}  
    & 1 & 0.793 & 0.791 & 0.796 & 0.785 & 0.774 & 0.728 \\
    & 2 & 0.802 & 0.793 & 0.798 & 0.793 & 0.788 & -- \\
    & 3 & $\bf{0.806}$ & 0.794 & 0.792 & 0.790 & 0.782 & -- \\
    & 4 & 0.795 & 0.790 & 0.798 & 0.783 & 0.792 & -- \\
    \midrule
    \multirow{4}{*}{PR-AUC}  
    & 1 & 0.814 & 0.814 & 0.816 & 0.810 & 0.800 & 0.637 \\
    & 2 & 0.823 & 0.814 & 0.821 & 0.814 & 0.810 & -- \\
    & 3 & $\bf{0.831}$ & 0.816 & 0.815 & 0.810 & 0.801 & -- \\
    & 4 & 0.817 & 0.812 & 0.824 & 0.801 & 0.812 & -- \\
    \midrule
    \multirow{4}{*}{ROC-AUC}  
    & 1 & 0.860 & 0.857 & 0.863 & 0.850 & 0.844 & $\bf{0.932}$ \\
    & 2 & 0.864 & 0.860 & 0.858 & 0.860 & 0.856 & -- \\
    & 3 & 0.859 & 0.860 & 0.855 & 0.859 & 0.858 & -- \\
    & 4 & 0.859 & 0.856 & 0.855 & 0.860 & 0.861 & -- \\
    \bottomrule
  \end{tabular}
\end{table*}

As a baseline, we used the cosine similarity-based segmentation method, which achieved the highest accuracy in the experiments reported in our previous work~\cite{2024_Y.Jin}.
This unsupervised approach computes the cosine similarity between the two items adjacent to each candidate segmentation point.
If the similarity exceeded a predefined threshold, the point was considered as a segmentation point.

\subsection{Effectiveness of the Proposed Method (RQ1)}
As summarized in Table~\ref{tab:performance}, the proposed method consistently outperformed the baseline across all supervised models with respect to the F1-score and PR-AUC.
This performance advantage holds for different window sizes, indicating the robustness of the proposed method.
In particular, increasing the window size beyond one improved the segmentation performance in most metrics.
LightGBM served as a representative example, achieving the highest overall scores.
For instance, its F1-score improved from 0.793 to 0.806 and its PR-AUC from 0.814 to 0.831, when the window size increased from one to three, demonstrating the benefit of incorporating a broader context.

In contrast, the baseline achieved the highest ROC-AUC of 0.932 at a window size of one, suggesting that the cosine similarity metric is highly effective in distinguishing positive and negative samples.
However, the F1-score and PR-AUC were lower, indicating that it might be less reliable for precise session segmentation than supervised methods.
A possible reason for the lower ROC-AUC of the proposed method is that hyperparameter tuning was performed with the F1-score, which emphasizes recall.
This likely caused the model to favor positive predictions, thereby increasing the number of false positives.
Given the class imbalance in the test samples (10.82\% positive vs. 89.18\% negative), this bias may have degraded performance on the negative class, reducing the ROC-AUC.

In summary, these results support RQ1, showing that the proposed method outperforms the baseline in the F1-score and PR-AUC, and that it benefits from a broader contextual window.

\subsection{Best Supervised Learning Model (RQ2)}

\subsubsection{Model Performance Comparison and Insights}
As summarized in Table~\ref{tab:performance}, LightGBM achieved the highest performance, with an F1-score of 0.806 and a PR-AUC of 0.831 at a window size of three, the best results among all the models.
These results suggest that LightGBM effectively identifies session boundaries while maintaining a balanced trade-off between precision and recall.
Futhermore, Table~\ref{tab:performance} reveals several notable trends across models and metrics.
The tree-based models (LightGBM, XGBoost, and CatBoost) consistently outperformed the linear models (SVM and LR) with respect to the F1-score and PR-AUC, indicating their advantage in capturing complex feature interactions.

Another key observation concerns the window size $w$.
Several models improved the performance as $w$ increased from one to two or three, but some showed a performance drop at $w = 3$ or $w = 4$.  
This suggests that while incorporating contextual information is helpful, excessive context or long-range dependencies may introduce noise and degrade segmentation accuracy.

In summary, LightGBM proved the most effective model under the proposed method, offering the best balance across the evaluation metrics and providing a clear answer to RQ2.

\begin{figure*}[t]
  \centering
  \begin{subfigure}{0.32\linewidth}
    \centering
    \includegraphics[width=\linewidth]{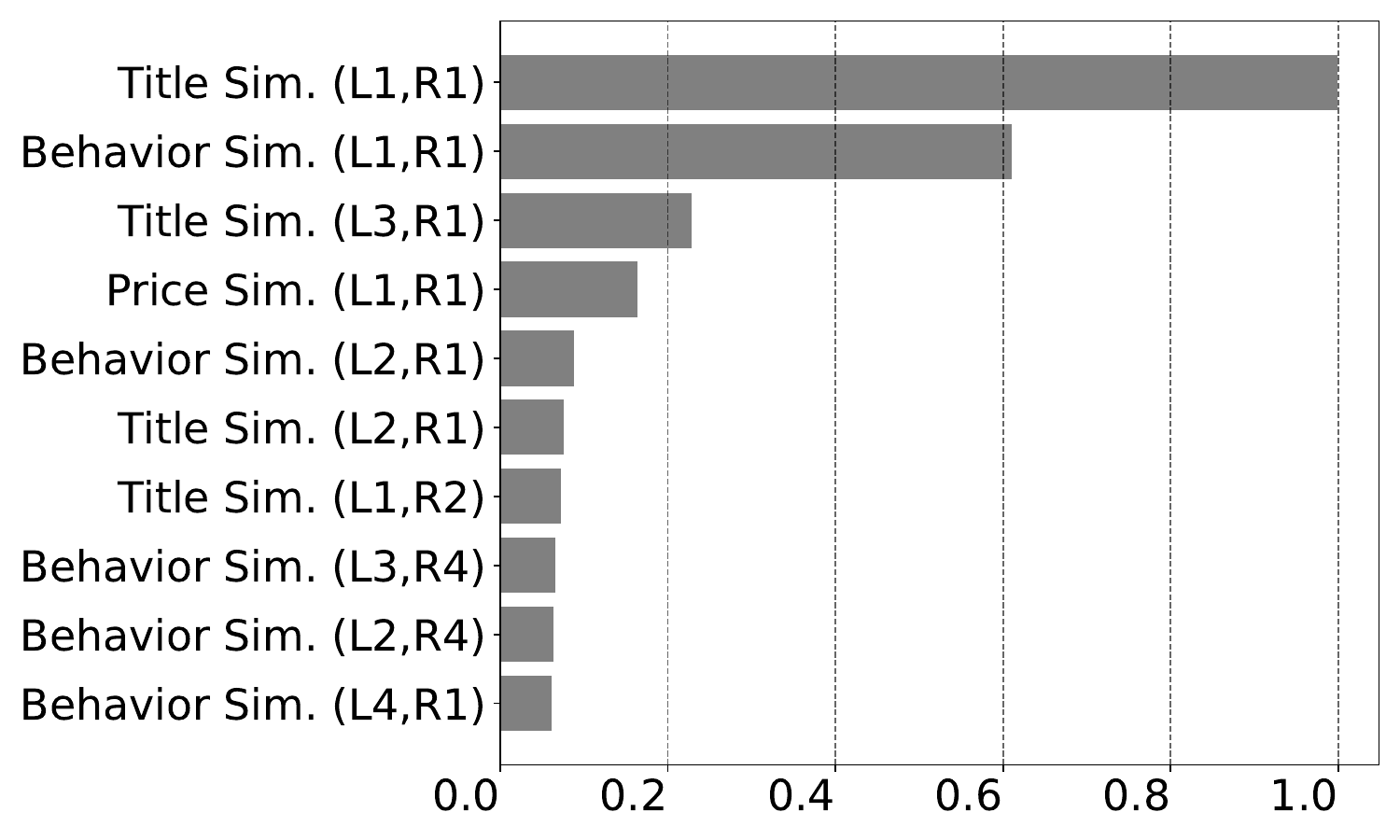}
    \caption{LightGBM importance}
    \label{fig:lgbm_importance}
  \end{subfigure}
  \begin{subfigure}{0.32\linewidth}
    \centering
    \includegraphics[width=\linewidth]{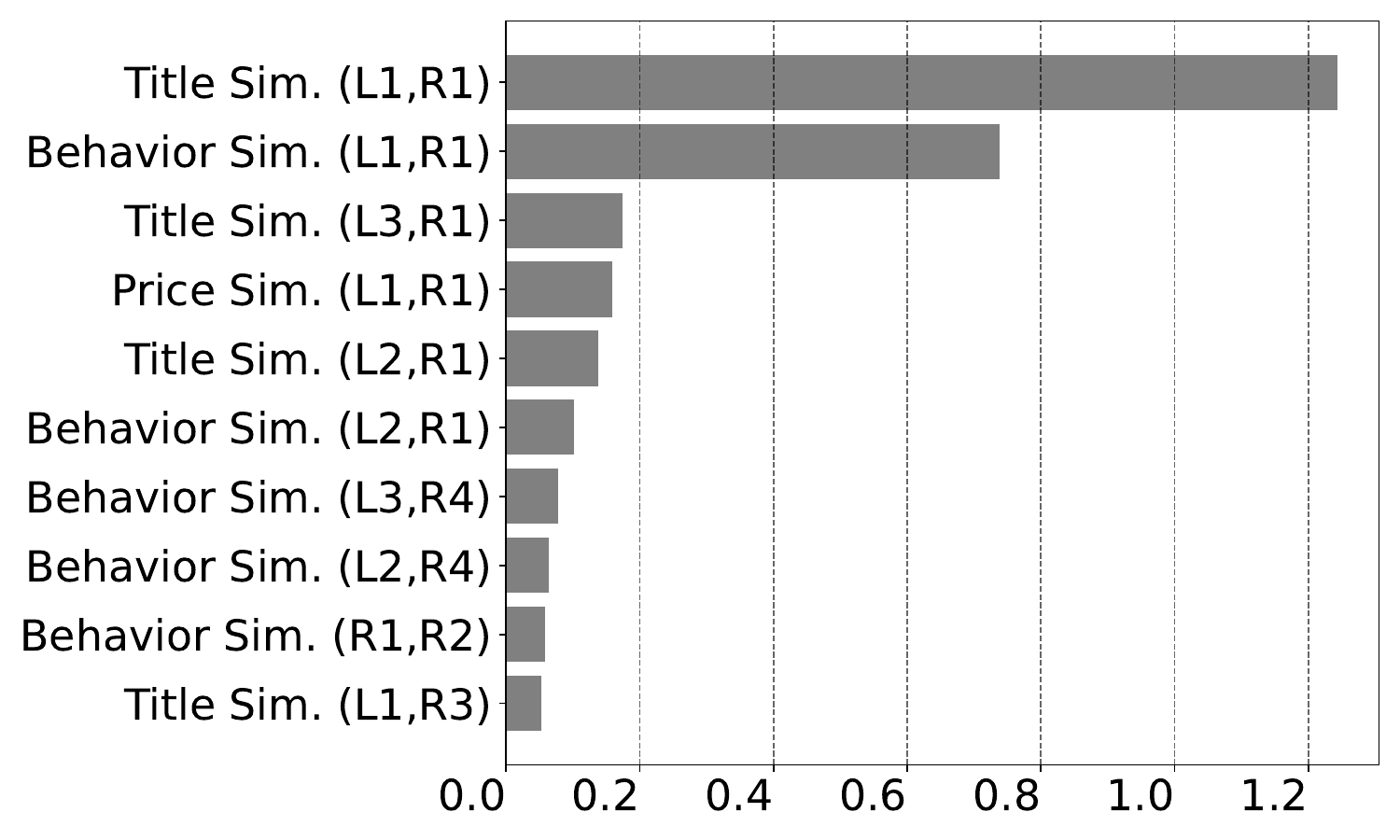}
    \caption{XGBoost Importance}
    \label{fig:xgboost_importance}
  \end{subfigure}
  \begin{subfigure}{0.32\linewidth}
    \centering
    \includegraphics[width=\linewidth]{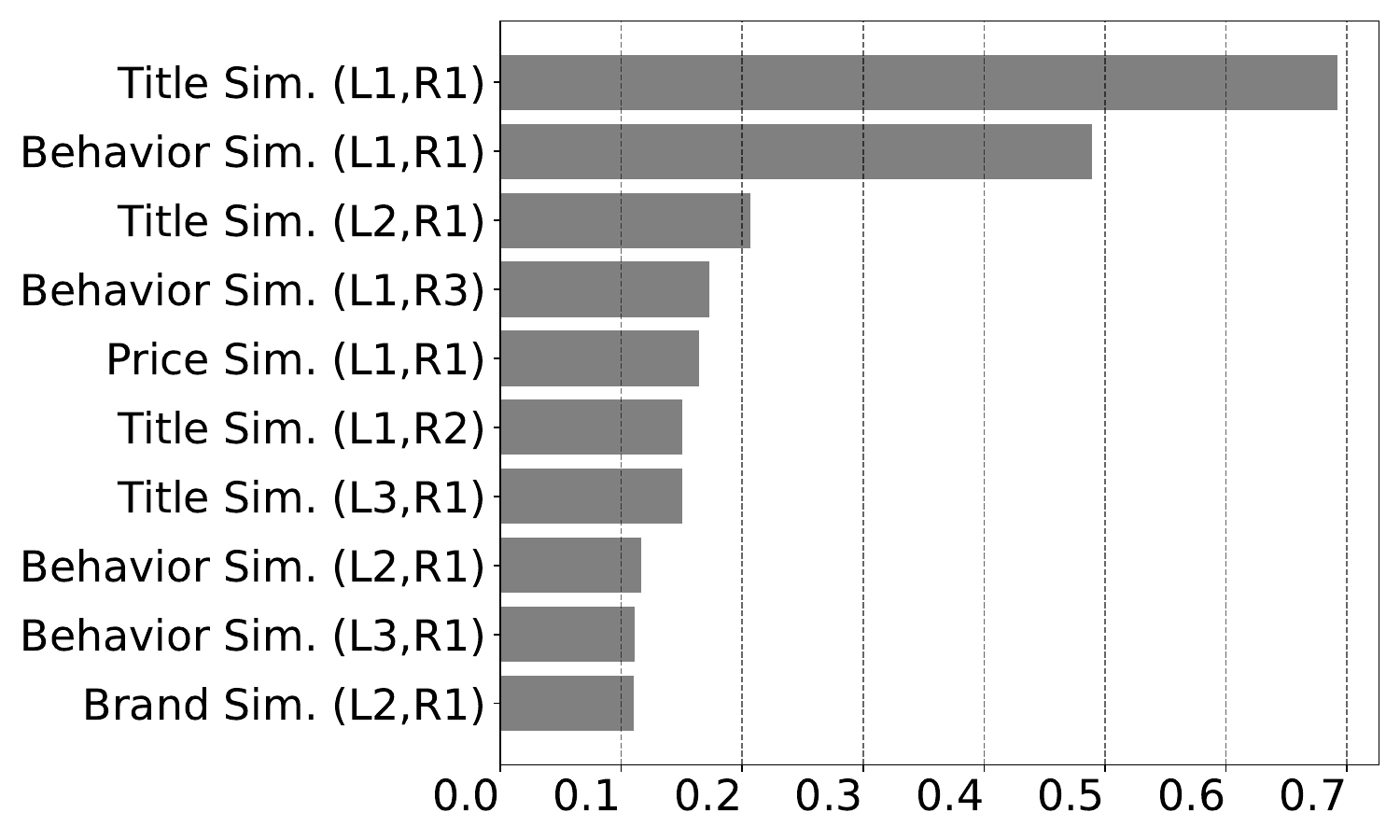}
    \caption{CatBoost importance}
    \label{fig:catboost_importance}
  \end{subfigure}
  
  \begin{subfigure}{0.32\linewidth}
    \centering
    \includegraphics[width=\linewidth]{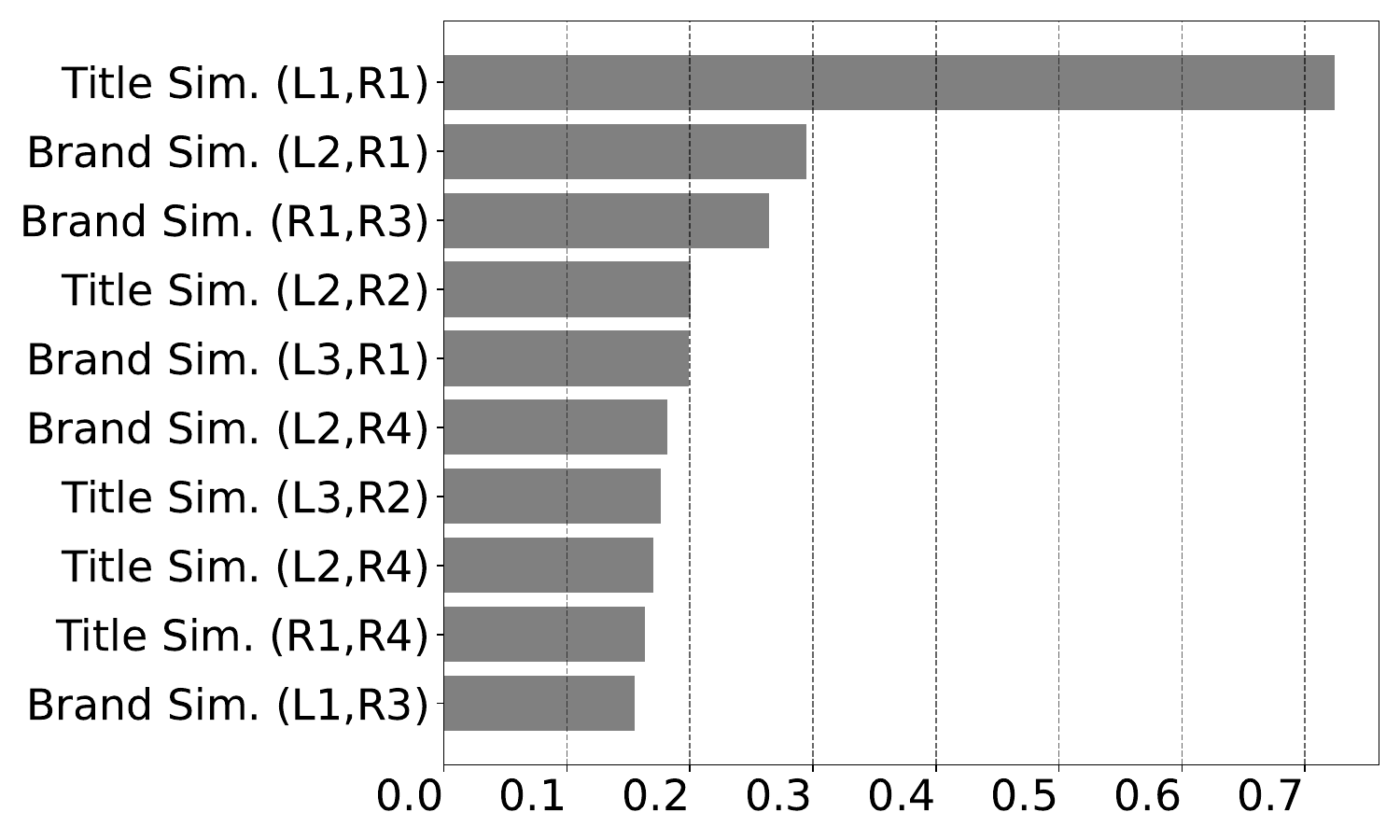}
    \caption{SVM Importance}
    \label{fig:svm_importance}
  \end{subfigure}
  \begin{subfigure}{0.32\linewidth}
    \centering
    \includegraphics[width=\linewidth]{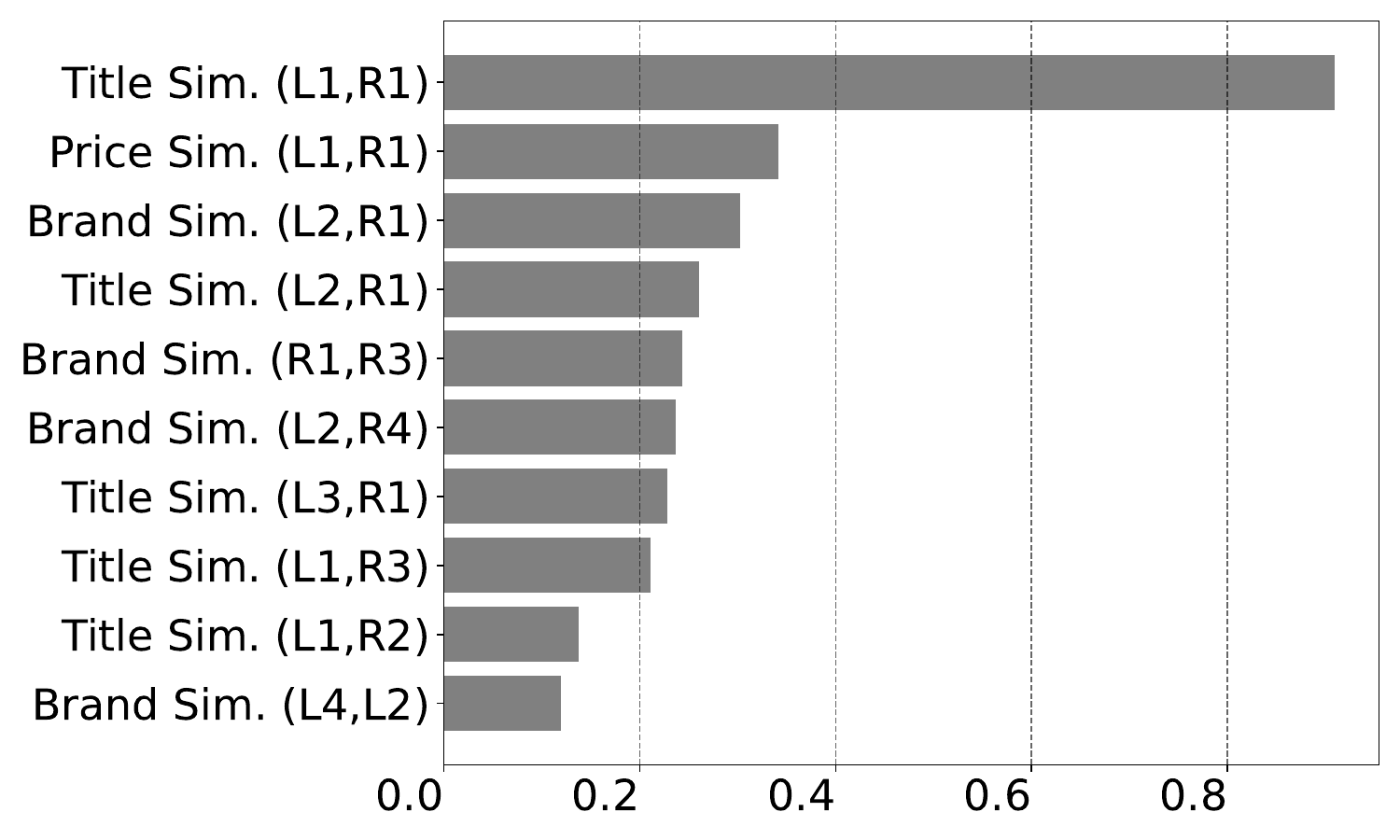}
    \caption{LR importance}
    \label{fig:lr_importance}
  \end{subfigure}
  \caption{Feature importance comparison across models}
  \label{fig:importance}
\end{figure*}

\subsubsection{Feature Importance Analysis}
We assessed the feature contributions to session segmentation using SHapley Additive exPlanations (SHAP) values, with the window size set to four.
Although the best performance was achieved with a window size of three, we used four to examine feature interactions in a broader context.
Feature importance was measured as the average magnitude of the SHAP values across test samples.
For clarity, feature pairs are denoted as $(L_{i}, R_{j})$, where $L_{i}$ and $R_{j}$ indicate the $i$-th item to the left and the $j$-th to the right of a potential segmentation point, respectively.
For instance, $(L_{1}, R_{1})$ represents the pair between the last item before and the first item after the segmentation point.

The results are shown in Fig.~\ref{fig:importance}.
Based on Figs.~\ref{fig:lgbm_importance}, \ref{fig:xgboost_importance}, and \ref{fig:catboost_importance}, item title similarity exhibits the strongest impact on predictions across all models.
Tree-based models (LightGBM, XGBoost, and CatBoost) rely heavily on behavior embeddings, in addition to title embeddings. 
By contrast, SVM and LR emphasize not only title embeddings but also brand similarity.

\section{Conclusion}
In this study, we proposed a supervised session segmentation method using similarity-based features derived from item embeddings and additional attributes.
The method evaluates each candidate segmentation point using features computed within a fixed-size window, capturing the contextual relationships between items.
We trained several classifiers, including LightGBM, XGBoost, CatBoost, SVM, and LR, and evaluated their performance on a manually annotated dataset constructed for this study.
Segmentation accuracy was assessed using the F1-score, ROC-AUC, and PR-AUC as the evaluation metrics.

The experimental results showed that the LightGBM-based model achieved the highest performance, with an F1-score of 0.806 and a PR-AUC of 0.831.
Compared with the baseline using cosine similarity with Item2Vec embeddings, the proposed method achieved improvements of 10.71\% in the F1-score and 30.61\% in the PR-AUC, confirming the effectiveness of incorporating contextual similarity features and supervised learning.
These findings suggest that the proposed method can serve as a reliable preprocessing step for downstream tasks, including session-based recommendation, user modeling, and behavioral pattern mining.
Moreover, the modular design allows for easy integration with different types of embeddings and additional item-level metadata.

Despite these promising results, several limitations still remain.
First, the method depends on manually annotated segmentation labels, which are costly and time-consuming to obtain at scale.
Second, because the dataset did not include user identifiers, the segmentation model applied a global criterion across all sessions, limiting its ability to capture personalized session boundaries and learn user-specific behavioral patterns.

In future work, we plan to explore semi-supervised and weakly supervised methods to reduce the dependence on manual annotation.
In addition, we intend to evaluate the practical impact of improved segmentation on the downstream tasks, such as click prediction and personalized recommendations.

\section*{Funding}
This work was supported by JSPS KAKENHI Grant Numbers JP23K21724 and JP24K21410.




\bibliographystyle{elsarticle-num} 
\bibliography{reference}






\end{document}